\documentclass[a4paper,fleqn,usenatbib]{mnras}

\usepackage{newtxtext,newtxmath}

\usepackage[T1]{fontenc}
\usepackage{ae,aecompl}

\usepackage{graphicx}	
\usepackage{amsmath}	
\usepackage{amssymb}	

\title[The aspherical explosion of SN 2017gmr]{The aspherical explosion of the Type IIP SN 2017gmr \thanks{Based on observations made with ESO Telescopes ant the Paranal Observatory under Program ID 099.D-0543.}}

\author[T. Nagao et al.]{
T. Nagao,$^{1,2}$\thanks{E-mail: Takashi.Nagao@eso.org}
A. Cikota,$^{3}$
F. Patat,$^{1}$
S. Taubenberger,$^{4}$
M. Bulla,$^{5}$
T. Faran,$^{6}$
\newauthor
D. J. Sand,$^{7}$
S. Valenti,$^{8}$
J. E. Andrews,$^{7}$
and D. E. Reichart$^{9}$
\\
$^{1}$European Southern Observatory, Karl-Schwarzschild-Str. 2, 85748 Garching b. M\"{u}nchen, Germany\\
$^{2}$Department of astronomy, Kyoto University, Kitashirakawa-Oiwake-cho,Sakyo-ku, Kyoto 606-8502, Japan\\
$^{3}$Physics Division, Lawrence Berkeley National Laboratory, 1 Cyclotron Road, Berkeley, CA 94720, USA\\
$^{4}$Max-Planck-Institut f\"{u}r Astrophysik, Karl-Schwarzschild-Str 1, 85748 Garching b. M\"{u}nchen, Germany\\
$^{5}$The Oskar Klein Centre, Physics Department, Stockholm University, SE-106 91 Stockholm, Sweden\\
$^{6}$Racah Institute of Physics, The Hebrew University of Jerusalem, Jerusalem 91904, Israel\\
$^{7}$Department of Astronomy and Steward Observatory, University of Arizona, 933 N. Cherry Avenue, Tucson, AZ 85719, USA\\
$^{8}$Department of Physics, University of California, Davis, CA 95616, USA\\
$^{9}$Department of Physics and Astronomy, University of North Carolina at Chapel Hill, Chapel Hill, NC 27599, USA
}

\date{Accepted XXX. Received YYY; in original form ZZZ}

\pubyear{2019}

\begin{document}
\label{firstpage}
\pagerange{\pageref{firstpage}--\pageref{lastpage}}
\maketitle

\begin{abstract}
Type IIP supernovae (SNe IIP), which represent the most common class of core-collapse (CC) SNe, show a rapid increase in continuum polarization just after entering the tail phase. This feature can be explained by a highly asymmetric helium core, which is exposed when the hydrogen envelope becomes transparent. Here we report the case of a SN IIP (SN~2017gmr) that shows an unusually early rise of the polarization, $\gtrsim 30$ days before the start of the tail phase. This implies that SN~2017gmr is an SN IIP that has very extended asphericity. The asymmetries are not confined to the helium core, but reach out to a significant part of the outer hydrogen envelope, hence clearly indicating a marked intrinsic diversity in the aspherical structure of CC explosions. These observations provide new constraints on the explosion mechanism, where viable models must be able to produce such extended deviations from spherical symmetry, and account for the observed geometrical diversity.
\end{abstract}




\section{Introduction}
Unveiling the explosion mechanism of core-collapse supernovae (CC SNe) is the key for understanding the chemical enrichment of galaxies and the induction of star formation \citep[e.g.,][]{Janka2012}. Notwithstanding the progress made in understanding CC SNe, many questions about the mechanisms that lead to the explosion of massive stars remain unanswered \citep[e.g.,][]{Janka2012}. The most promising scenario is the so-called neutrino-driven mechanism, where a star explodes due to neutrino heating from a proto-neutron star \citep[e.g.,][]{Janka2007}. This model reproduces some observed properties such as energy and chemical yields. However, it appears to work well only for low-mass progenitor stars \citep[i.e., stars whose mass is between $\sim 8$ M$_{\odot}$ and $\sim 10$ M$_{\odot}$; e.g.,][]{Janka2007}. For more massive stars, the situation is less satisfactory. It has been shown that one-dimensional simulations cannot lead to SN explosions \citep[e.g.,][]{Rampp2000, Liebendorfer2001, Thompson2003, Sumiyoshi2005}. The multi-dimensional effect is believed to be essential in the SN explosion mechanism. In fact, some multi-dimensional simulations reported success in launching the explosion, even though the calculated energy does not reach the values estimated from observations \citep[$\sim 10^{51}$ erg; e.g.,][]{Buras2006, Marek2009, Takiwaki2012, Hanke2013, Melson2015}. In order to solve this problem, it is important to investigate the information on the explosion structure of SNe given by observations. Polarimetry provides the most reliable tool for getting insights on the explosion geometry, which is a crucial aspect in this problem and can hardly be investigated by any other technique.

Type IIP SNe (SNe IIP) represent the most common class of CC SNe \citep[$\sim 50$ per cent of all core-collapse events; e.g.,][]{Li2011}. SNe IIP show constant magnitude in the R/I bands until $\sim 100$ days after the explosion (the plateau phase), which is followed by an exponential decline (the tail phase) after a sudden drop \citep[see the gray line in Fig.~1; e.g.,][]{Anderson2014, Sanders2015, Valenti2016}. They also display a rapid increase in the continuum polarization ($\sim 1.0$ per cent) just after the end of the plateau phase, following a generally low polarization level ($\sim 0.1$ per cent) during the plateau phase \citep[see the coloured dots in Fig.~1; e.g.,][]{Wang2008}. This behaviour can be explained in terms of a highly asymmetric helium core that is revealed when the hydrogen envelope has become transparent. Although only a few SNe IIP have been observed with polarimetry in the tail phase, this is regarded as supporting evidence for an asymmetric explosion \citep[see][for a review]{Wang2008}. Here we report an extensive polarimetric study aimed at determining the explosion geometry of a CC SN. For this purpose we have observed a nearby Type IIP SN (SN~2017gmr) from the plateau phase to the tail phase.

\begin{table*}
      \caption[]{Log of the observations of SN 2017gmr.}
\scalebox{0.89}[0.89]{$\displaystyle
         \begin{array}{lcccccccc}
            \hline
            \noalign{\smallskip}
            \rm{Date} & \rm{MJD} & \rm{Phase}^{a} & \rm{Days\; from \;detection}^{b} & \rm{Airmass} & \rm{Exp. \;time} & \rm{Pol.\;degree} & \rm{Pol. \;angle} & \rm{Obs. \;mode}\\
            (\rm{UT}) & (\rm{days}) & (\rm{days}) & (\rm{days}) & (\rm{average}) & (\rm{s}) & (\rm{per\;cent}) & (\rm{degrees}) & \\
            \noalign{\smallskip}
            \hline\hline
            \noalign{\smallskip}      
            2017-10-19.60 & 58045.60 & -54.4 & +45.40 & 1.2 & 4 \times 3110 & 0.18 \pm 0.02 & 87.3 \pm 4.7 & \rm{PMOS}\\
            \noalign{\smallskip} \hline \noalign{\smallskip}
            2017-11-10.24 & 58067.24 & -32.76 & +67.04 & 1.2 & 4 \times 660 & 0.69 \pm 0.04 & 92.5 \pm 1.7 & \rm{PMOS}\\
            \noalign{\smallskip} \hline \noalign{\smallskip}
            2017-11-24.24 & 58081.24 & -18.76 & +81.04 & 1.4 & 4 \times 120 & 1.08 \pm 0.16 & 105.1 \pm 5.5 & \rm{IPOL}\\
            \noalign{\smallskip} \hline \noalign{\smallskip}
            2017-12-12.86 & 58099.86 & -0.14 & +99.66 & 1.2 & 4 \times 2640 & 0.89 \pm 0.03 & 101.5 \pm 0.9 & \rm{PMOS}\\
            \noalign{\smallskip} \hline \noalign{\smallskip}
            2017-12-17.22 & 58104.22 & +4.22 & +104.02 & 1.8 & 4 \times 120 & 1.03 \pm 0.12 & 96.6  \pm 4.2 & \rm{IPOL}\\
            \noalign{\smallskip} \hline \noalign{\smallskip}
            2017-12-21.08 & 58108.08 & +8.08 & +107.88 & 1.1 & 4 \times 240 & 0.77 \pm 0.09 & 102.6 \pm 5.1 & \rm{IPOL}\\ 
            \noalign{\smallskip} \hline \noalign{\smallskip}
            2017-12-21.76 & 58108.76 & +8.76 & +108.56 & 1.2 & 4 \times 3960 & 0.65 \pm 0.03 & 100.4 \pm 1.4 & \rm{PMOS}\\
            \noalign{\smallskip} \hline \noalign{\smallskip}
            2018-01-14.09 & 58132.09 & +32.09 & +131.89 & 1.3 & 4 \times 280 & 0.65 \pm 0.11 & 91.1  \pm 7.0 & \rm{IPOL}\\ 
            \noalign{\smallskip} \hline \noalign{\smallskip}
            2018-01-16.93 & 58134.93 & +34.93 & +134.73 & 1.3 & 4 \times 4620 & 0.37 \pm 0.04 & 105.9 \pm 6.4 & \rm{PMOS}\\
            \noalign{\smallskip} \hline \noalign{\smallskip}
            2018-02-11.03 & 58160.03 & +60.03 & +159.83 & 1.4 & 4 \times 280 & 0.00 \pm 0.15 & 41.4  \pm 12.2 & \rm{IPOL}\\
            \noalign{\smallskip}
            \hline
         \end{array}
         $}
         \begin{minipage}{.88\hsize}
        \smallskip
        Notes. ${}^{a}$Relative to $t_{0}=58100$ (MJD), which is the timing of the end of the plateau phase. ${}^{b}$Relative to $t=58000.20$ (MJD), which is the timing of the first detection and almost the explosion time. The non-detection was reported 1.97 days before the first detection. For the sake of increasing the signal-to-noise ratio, we combine the spectropolarimetric data into five groups, where we have checked the consistency of the spectra within each group.
        \end{minipage}
\end{table*}

\begin{figure}
  \includegraphics[width=\columnwidth]{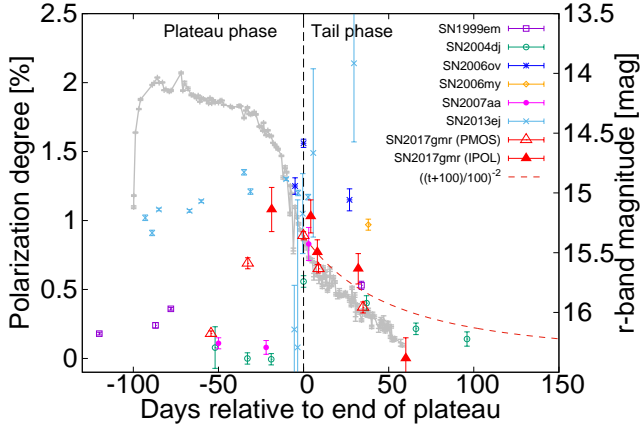}
    \caption{Intrinsic continuum polarization of SN~2017gmr, compared with other SNe IIP. The gray crosses connected by a line show the SDSS $r$-band light curve of SN~2017gmr taken by the PROMPT5 telescopes. The black vertical dashed line indicates the timing of the end of the plateau phase, $t_{0}=58100$ (MJD). The continuum polarization estimated from the spectro- and imaging-polarimetric observations is shown as red open triangles and red filled triangles, respectively. The red dashed line represents the expected decline in polarization ($P=P_{0}((t+100)/100)^{-2}$, where $P_{0}=0.89$ per cent and the plateau duration is $100$ days) due to the effects of decreasing optical depth in the expanding ejecta. The data for SNe 1999em, 2004dj, 2006ov, 2006my, 2007aa and 2013ej (purple squares, green circles, blue stars, orange diamonds, magenta dots and cyan crosses, respectively) are taken from \citet[][]{Leonard2001, Leonard2006, Chornock2010, Kumar2016, Mauerhan2017}. The time of the end of the plateau is 51607.0, 53290.4, 54094.5, 54084.0, 54227.3 and 56602.0 (MJD) for SNe 1999em, 2004dj, 2006ov, 2006my, 2007aa and 2013ej, respectively.}
\end{figure}

\section{Observations and data reduction}
We have conducted spectropolarimetric and imaging-polarimetric observations of the Type IIP SN 2017gmr, using the FOcal Reducer/low-dispersion Spectrograph 2 (hererafter FORS2) mounted at the Cassegrain focus of the Very Large Telescope (VLT) UT1 (Antu) telescope in Chile. SN~2017gmr was discovered in NGC 0988 during the ongoing DLT40 one-day cadence supernova search \citep[][]{Tartaglia2018} on 4.20 September 2017 UT \citep[58000.20 MJD;][]{Valenti2017}, located at $z=0.005075$ and receding with a velocity $v_{\rm{gal}}=1517.5$ km s$^{-1}$ \citep[][]{Meyer2004}. A few days later, the object was classified as a CC SN \citep[][]{Pursimo2017}. The object was not detected on 2.23 September 2017 UT (57998.23 MJD), i.e. about two days before the discovery \citep[][]{Valenti2017}. We have conducted spectropolarimetric (PMOS) observations of SN~2017gmr from $\sim 50$ to $\sim 140$ days after the discovery (i.e., the explosion), covering the plateau and the transition to the tail phase. In addition, when the object had faded below the spectroscopic threshold, we performed imaging polarimetry (IPOL) through the narrow band filter FILT\_815\_13, with a central wavelength of 815 nm (811 nm in the restframe of the host galaxy), from $\sim 80$ days after the discovery to epochs extending well into the tail phase. This filter covers a continuum-dominated wavelength range and was specifically selected to avoid any contamination from spectral lines. The observations log is given in Table~1, where the phase is counted from the end of the plateau phase, $t_{0}$. This was determined, using the DLT40 $r$-band light curve, as the epoch when the luminosity decline begins to follow the radioactive exponential tail (see Fig.~1): $t_{0}=58100$ (MJD).

The SN was observed at several epochs, using the dual-beam polarimeter FORS2 in PMOS and IPOL modes (Table~1). For the spectropolarimetric observations, the spectrum produced by a grism is split by the Wollaston prism into two beams with orthogonal polarization directions: the ordinary (o) and extraordinary (e) beams. The beam splitter is coupled to a half-wave retarder plate (HWP), which allows the measurement of the mean electric field intensity along different angles on the plane of the sky. For our observations we adopted the optimal angle set $0^{\circ}$, $22.5^{\circ}$, $45^{\circ}$ and $67.5^{\circ}$ \citep[see][for more details]{Patat2006}. The HWP angle is measured between the acceptance axis of the ordinary beam of the Wollaston prism (aligned to the north-south direction) and the fast axis of the retarder plate. As a dispersive element we used the low-resolution G300V grism coupled to a $1.1$ arcsec slit, and it can give a spectral range $3800-9200$ \AA, a dispersion of $\sim 3.3$ \AA\; pixel$^{-1}$ and a resolution of $11.6$ \AA\; ($FWHM$) at $5580$ \AA. For the imaging-polarimetric observations, the same instrumental setup was used, with a narrow band filter (FILT\_815\_13) replacing the grism in the optical path.

The data were reduced by the standard methods as described, e.g., in \citet[][]{Patat2006} with IRAF\footnote{IRAF is distributed by the National Optical Astronomy Observatories, which are operated by the Association of Universities for Research in Astronomy, Inc., under cooperative agreement with the National Science Foundation.}. The ordinary and extraordinary beams of the spectropolarimetric data were extracted by the PyRAF apextract.apall task, using a fixed aperture size of $10$ pixels and then separately rebinned to 50 Angstrom bins for improving the signal-to-noise ratio. The HWP zeropoint angle chromatism was corrected using tabulated data \citep[][]{Jehin2005}.  The wavelength scale was corrected to the rest-frame using the galaxy redshift ($z=0.005075$) following the interstellar polarization (ISP) subtraction. The flux in the ordinary and extraordinary beams of the imaging polarimetry was measured with a fixed aperture radius of $1.5 \times \rm{FWHM}$, followed by the ISP subtraction. The polarization bias in the spectro- and imaging-polarimetric data were subtracted using the standard method in \citet[][]{Wang1997}.

The $r$-band light curve was obtained in the context of the DLT40 supernova search. The images of SN~2017gmr were taken by the PROMPT5 0.41m telescope located at CTIO and an identical PROMPT telescope at Meckering Observatory, Australia, using an 'Open' and 'Clear' filter, respectively.  Instrumental magnitudes were converted to an approximate $r$-band magnitude using the APASS catalog.

\section{Reddening and interstellar polarization} The Galactic reddening along the line of sight to SN~2017gmr is $E(B-V)=0.024$ \citep[][]{Schlafly2011}. The empirical relation by \citet[][]{Serkowski1975} indicates that the Galactic ISP should be lower than $\sim 0.2$ per cent. The extinction within the host galaxy is estimated to be relatively high \citep[][]{Elias-Rosa2017}, $E(B-V)=0.23$. Thus, in the extreme case when all the dust grains along the line of sight are aligned in the same direction, the total ISP can reach $\sim 2$ per cent. There are several ways to estimate the ISP component from spectro-polarimetry \citep[e.g.,][]{Trammell1993, Wang1997, Tran1997, Leonard2000, Wang2001, Howell2001, Wang2004, Leonard2005, Chornock2006, Patat2012, Reilly2017}, using, e.g., the assumptions (1) that the emission peaks of the lines with P-Cygni profiles have no intrinsic polarization, and (2) that spectra were taken at a sufficiently late phase that the signal inherent to the SN is completely gone and one is left with the pure ISP component. Since very late spectro-polarimetry is not available for this event, we adopted the first method. This is based on the following reasonings. For optically thick lines, multiple scattering processes tend to weaken the geometrical imprint carried by the photospheric radiation and hence depolarize emission lines. In addition, collisional redistribution of the atomic state of excited atoms during absorption and reemission processes in a line also tend to extinguish the polarization information \citep[e.g.,][]{Hoflich1996, Kasen2006}.
This implies that any polarization signal measured at emission line peaks would give a direct estimate of the ISP.

Figure 2a shows polarimetric spectra at $t=58099.86$ (a phase of -0.14 days, which is the closest epoch to the polarization peak) before the ISP subtraction. Indeed, the peaks of the most prominent emission lines (e.g., H$\alpha$, Ca II triplet) display non-null polarization, with an angle that differs from that measured in the continuum range. This implies the existence of a non-intrinsic component with a different position angle, i.e. the ISP. The polarization angle across the emission lines is around 30 degrees, which we assume is the ISP angle. We have derived the ISP wavelength dependency by fitting the polarimetric data of all epochs with the classical Serkowski function \citep[][]{Serkowski1975}: $P(\lambda) = P_{\rm{max}} \exp \left[ -K \ln^{2} \left( \lambda_{\rm{max}}/\lambda \right) \right]$,
restricting the fit to wavelength windows where the polarization angle is between 20 and 40 degrees. The best-fit values for the parameters are $P_{\rm{max}}=0.4$ per cent, $\lambda_{\rm{max}}=4900$ \AA\; and $K=1.1$; thus $Q_{ISP}$ at $\lambda=4900$ \AA\; is $0.20$ per cent and $U_{ISP}$ at $\lambda=4900$ \AA\; is $0.35$ per cent (see Fig. 2a).

\begin{figure*}
  \includegraphics[width=0.9\columnwidth]{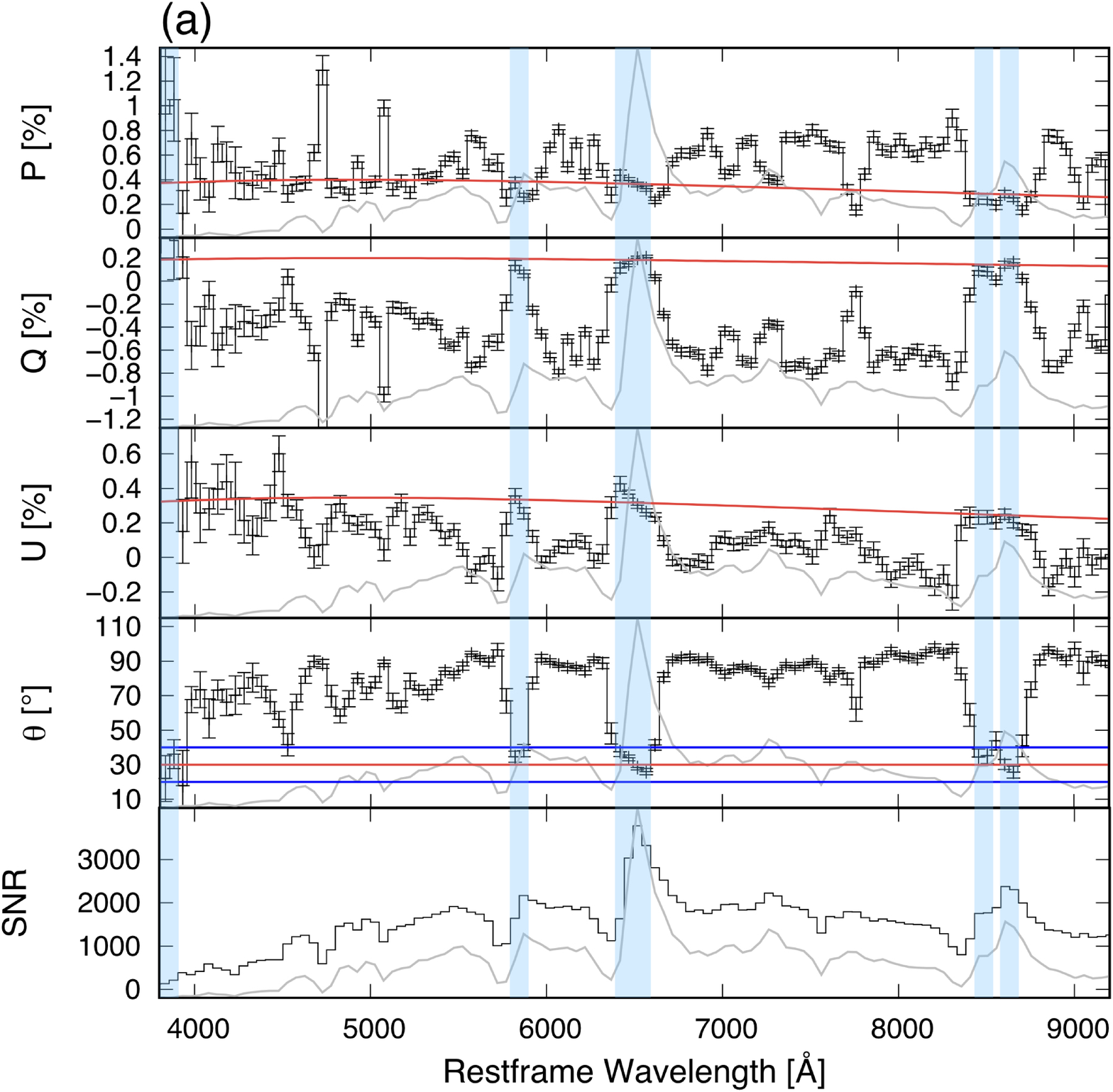}
  \includegraphics[width=0.9\columnwidth]{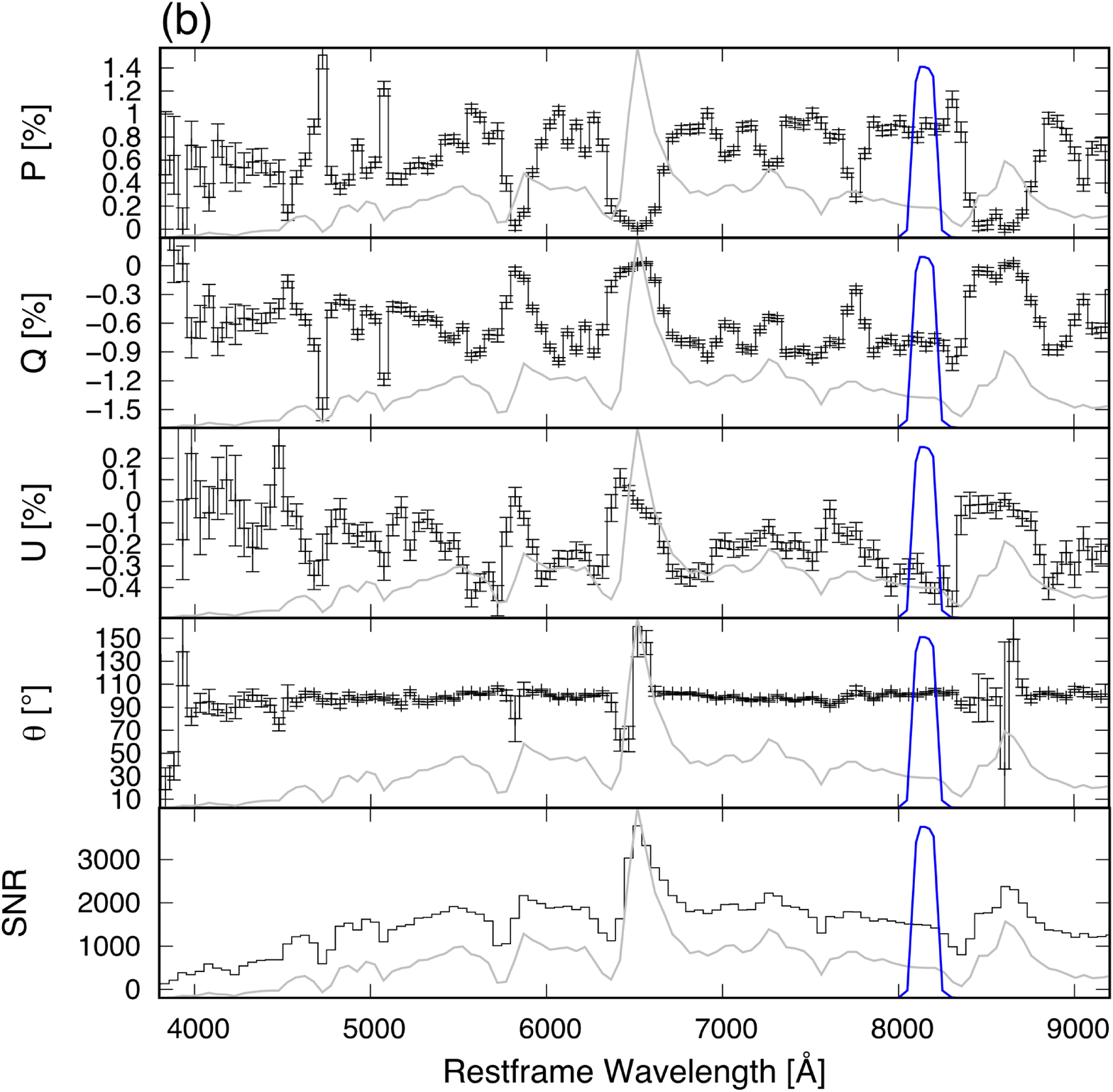}
    \caption{Polarization spectra of SN~2017gmr before and after the ISP subtraction. (a) Total polarization $P$, Stokes parameters $Q$ and $U$, polarization angle $\theta$, and signal-to-noise ratio (SNR) for SN~2017gmr before the ISP subtraction at a phase of -0.14 days (black lines). The data is binned to $50$ \AA\; per point. The gray lines in the background of each plot are the unbinned total-flux spectra at the same epoch. The ISP is described by $P(\lambda) = P_{\rm{max}} \exp \left[ -K \ln^{2} \left( \lambda_{\rm{max}}/\lambda \right) \right]$, where $P_{\rm{max}}=0.4$, $\lambda_{\rm{max}}=4900$ \AA\; and $K=1.1$. The red line and blue lines in the polarization angle plot represent the assumed ISP angle and the adopted maximum and minimum polarization angle for determing the ISP, respectively. The blue hatching shows the adopted wavelength range for the ISP dominated components. (b) Same as (a), but after ISP subtraction. The blue lines in the plots trace the transmission curve of FILT\_815\_13.}
\end{figure*}

\section{Results \& Discussions}
Figure 2b shows polarization spectra after the ISP subtraction. The intrinsic continuum polarization is characterized by a single polarization angle ($\theta_{\rm{PA}} \sim 100$ degrees) and a constant, wavelength-independent polarization degree. It should be noted that the polarization in the bluer wavelength is heavily contaminated by the line polarization (e.g., line-blanketing by Fe at shorter wavelength than 5500 Angstrom; see Fig.~2). To study its time-evolution, we estimated the continuum polarization from the spectropolarimetric data by integrating the spectra across the pass-band of FILT\_815\_13 filter (see the blue lines in Fig. 2b), thus ensuring full consistency between imaging and spectroscopic measurements. The results are presented in Table~1 and Fig.~1. This shows that the polarization levels derived from the two different techniques are fairly consistent for similar epochs, which were deliberately obtained to allow a direct cross-check.
The time-evolution of the continuum polarization is presented in Fig.~1, which also shows a comparison with other SNe IIP. The behavior of SN~2017gmr during the decline phase of the polarization is similar to that observed for other SNe IIP, in particular to the well-observed SN IIP 2004dj \citep[the polarization degree is proportional to $t^{-2}$;][]{Leonard2006}. A closer look reveals that SNe 2006ov and 2006my show higher polarization degrees (which may be explained by viewing-angle effects), while SN~2013ej displays a largely different behavior. The big difference between SN~2017gmr and all other objects except SN~2013ej is the timing of the polarization rise. SN~2017gmr shows an early rise with a comparatively high degree of polarization already at $\sim 30$ days before the start of the tail phase, although SN~2013ej shows high polarization at much earlier epochs, starting at nearly the explosion epoch.

The early rise of the polarization can be explained by electron scattering in asymmetric distribution of photosphere and/or dust scattering in aspherically distributed circumstellar matter (CSM; the dust scattering model). In the dust scattering model, the light scattered by circumstellar dust accounts for the polarization, without requiring any inherent asphericity \citep[e.g.,][]{Wang1996, Nagao2017}. A method for testing this hypothesis has been recently proposed by \citet[][]{Nagao2018}, which is based on the wavelength dependence of the polarization. Since electron scattering is characterized by a gray opacity, the polarization signal generated by an aspherical photosphere is expected to be almost wavelength-independent. On the other hand, dust scattering processes strongly depend on wavelength, which generally lead to a higher degree of polarization at shorter wavelengths \citep[][]{Nagao2018}. Since our data show no wavelength dependence (see Fig.~2), we can reject the dust scattering model to explain the observed polarization at any epochs. Another important aspect is related to the early polarization rise. As shown in Fig.~1, the continuum polarization reaches $\sim 0.7$ per cent already at $\sim 30$ days before the end of the plateau phase. In the dust scattering model, the rise of the polarization corresponds to the drop of the SN luminosity \citep[][]{Nagao2017}. Thus, this scenario cannot be reconciled with our detection of a comparatively high degree of polarization well before the end of the plateau phase. Based on these two facts, we conclude that the main source of continuum polarization in SN~2017gmr is indeed an aspherical photosphere.

As an origin of the aspherical photosphere, there are the following possibilities: aspherical CSM interaction or aspherical explosion geometry \citep[e.g.,][]{Mauerhan2017}. In the CSM-interaction scenario by \citet[][see its Figure 16]{Mauerhan2017}, aspherical photosphere is created by additional heating from embedded interaction between SN ejecta and an aspherical CSM, so that an SN show high polarization without leaving any traces of CSM interaction in its optical light. Since the photosphere should be outside of the shock in this scenario, this scenario does not work after receding of the photosphere, i.e., after leaving the plateau phase. However, SN~2017gmr shows high polarization with a normal SN spectrum at a phase of -0.14 days (Fig.~1 and 2). In addition, the polarization decline of SN~2017gmr follows the $t^{-2}$ evolution, which can be explained by the effects of diminishing optical depth of optically thin and expanding ejecta. Thus, the aspherical photosphere should originate from aspherical explosion geometry. Finally, we conclude that the origin of the polarization in SN 2017gmr is aspherical explosion geometry. This identification of the origin of the polarization for SNe IIP is the first time ever, where \citet[][]{Nagao2018} plays important roles.

The early rise of the polarization implies that SN~2017gmr has a very extended aspherical geometry: asymmetries are not only present in the helium core, but they also extend to a significant part of the hydrogen envelope. The maximum polarization of $\sim 0.9$ per cent implies a significant departure from spherical symmetry of the inner region, with a minimum axis ratio of 1.2:1 in the electron-scattering atmosphere model by \citet[][]{Hoflich1991}. Since the minimum axis ratio of 1.2:1 corresponds to the case in which the SN is viewed on the equatorial plane, larger asphericities are required if the line of sight is closer to the symmetry axis of the ellipsoid. This fact provides new constraints on the explosion mechanism of CC SNe. Viable explosion models must be able to produce such extended asphericity, and account for the diversity observed in the explosion geometry of different SNe IIP. 

In addition, clarifying relations between the polarization properties (e.g., the peak polarization degree, the timing of the rise, the slope at the rising phase, the slope at the decline phase, etc) and the SN properties (e.g, peak luminosity, plateau length, ejecta velocity, metal abundance, etc.), i.e., linking the explosion asphericity to the explosion physics, is important to place stronger constraints on the explosion mechanism. 
For SN~2017gmr, the absolute magnitude in the $r$ band during the plateau phase ($M_{\rm{plateau}}(r)$) is $\sim -17.5$, which is relatively bright among SNe IIP. The time span between the first detection (almost coinciding with the last non-detection) and the end of the plateau is $\sim 100$ days, which is also rather long when compared to typical IIP event. SN~2017gmr stands out from the typical behavior of this class of objects also from the spectroscopic point of view, displaying significantly blue-shifted lines with larger expansion velocities \citep[see][for more detail observational features of SN~2017gmr]{Andrews2019}. This is illustrated in Fig.~3, which compares the H$\alpha$ line profile of SN~2017gmr to that of a prototypical event \citep[SN~1999em; $M_{\rm{plateau}}(R)\sim -16.4$;][]{Faran2014} and a luminous event \citep[SN~1996W; $M_{\rm{plateau}}(R)\sim -17.8$;][]{Inserra2013}. The resemblance between SN~2017gmr and SN~1996W is very pronounced, although we notice that SN~1996W was faster evolving. The figure also shows the line profiles for SNe 2004dj, 2006ov and 2007aa, which do not show signs of an extended aspherical explosion, but display a rise in the polarization around the transitional time from the plateau to tail phases (see Fig.~1).
The line of SN~2017gmr is more blue-shifted and wider than any of these objects, despite the fact that their spectra in Fig.~3 were taken at a similar or earlier epoch. As pointed out by \citet[][]{Dessart2011}, this blue-shift of the line might be due to the optical-depth effects by an aspherical explosion. Although we cannot exclude concurrent viewing-angle effects, this similarity and discrepancy of the line profiles can be interpreted as a sign that the extended asphericity seen in SN~2017gmr is related to a more energetic explosion. This might imply that SN~2017gmr is a jet-driven explosion \citep[e.g.,][]{Couch2009}. A firmer conclusion will have to wait for a larger polarimetric sample of SNe IIP.

\begin{figure}
  \includegraphics[width=0.9\columnwidth]{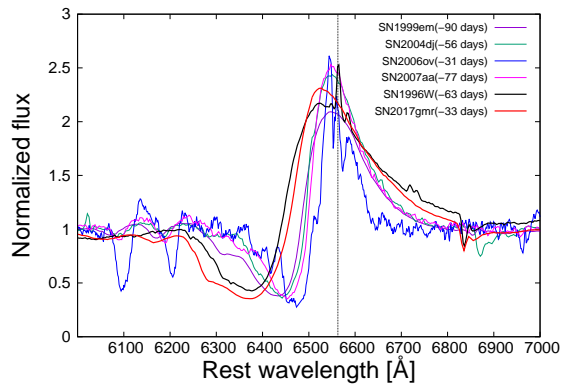}
    \caption{Comparison of H$\alpha$ lines in SN~2017gmr (red line) and other SNe IIP. The spectra are continuum-normalized, and shifted in wavelength to match the positions of the interstellar Na~I D lines. The data for SNe 1999em, 2004dj, 2006ov, 2007aa and 1996W (purple, green, blue, magenta and black lines, respectively) are taken from \citet[][]{Faran2014, Vinko2006, Hicken2017, Gutierrez2017, Inserra2013}, respectively. The numbers shown in the legends indicate the epoch from the end of the plateu phase (i.e. larger negative numbers indicate epochs closer to the explosion date). For SN~1996W, the time of the end of the plateau has been determined, assuming that SN~1996W has the same light curve as SN~2017gmr.}
\end{figure}

\vspace{\baselineskip}
\noindent
{\bf Acknowledgements.} We would like to thank K. Maeda, R. Ouchi and S. Benetti for useful discussions. This paper is based on observations made with ESO Telescopes at the Paranal Observatory under program ID 099.D-0543(A) and observations collected by DLT40 supernova search. The authors are grateful to ESO-Paranal staff for the support given during the service mode observations of SN~2017gmr. T.~N. is supported by Japan Society for the Promotion of Science (JSPS) Overseas Research Fellowship, and was supported by JSPS Overseas Challenge Program for Young Researchers. M.B. acknowledges support from the Swedish Research Council (Vetenskapsr\aa det) and the Swedish National Space Board. Research by DJS is supported by NSF grants AST-1821987 and 1821967. Research by SV is supported by NSF grants AST-1813176.





\begin{thebibliography}{99}
\bibitem[Anderson et al.(2014)]{Anderson2014} Anderson, J.~P., Gonz{\'a}lez-Gait{\'a}n, S., Hamuy, M., et al.\ 2014, \apj, 786, 67 
\bibitem[Buras et al.(2006)]{Buras2006} Buras, R., Rampp, M., Janka, H.-T., \& Kifonidis, K.\ 2006, \aap, 447, 1049 
\bibitem[Chornock et al.(2006)]{Chornock2006} Chornock, R., Filippenko, A.~V., Branch, D., et al.\ 2006, \pasp, 118, 722 
\bibitem[Chornock et al.(2010)]{Chornock2010} Chornock, R., Filippenko, A.~V., Li, W., \& Silverman, J.~M.\ 2010, \apj, 713, 1363 
\bibitem[Couch et al.(2009)]{Couch2009} Couch, S.~M., Wheeler, J.~C., \& Milosavljevi{\'c}, M.\ 2009, \apj, 696, 953
\bibitem[Dessart \& Hillier(2011)]{Dessart2011} Dessart, L., \& Hillier, D.~J.\ 2011, \mnras, 415, 3497 
\bibitem[Elias-Rosa et al.(2017)]{Elias-Rosa2017} Elias-Rosa, N., Pursimo, T., Korhonen, H., et al.\ 2017, The Astronomer's Telegram, 10746
\bibitem[Faran et al.(2014)]{Faran2014} Faran, T., Poznanski, D., Filippenko, A.~V., et al.\ 2014, \mnras, 442, 844 
\bibitem[Guti{\'e}rrez et al.(2017)]{Gutierrez2017} Guti{\'e}rrez, C.~P., Anderson, J.~P., Hamuy, M., et al.\ 2017, \apj, 850, 89 
\bibitem[Hanke et al.(2013)]{Hanke2013} Hanke, F., M{\"u}ller, B., Wongwathanarat, A., Marek, A., \& Janka, H.-T.\ 2013, \apj, 770, 66 
\bibitem[Hicken et al.(2017)]{Hicken2017} Hicken, M., Friedman, A.~S., Blondin, S., et al.\ 2017, \apjs, 233, 6 
\bibitem[H{\"o}flich(1991)]{Hoflich1991} H{\"o}flich, P.\ 1991, \aap, 246, 481 
\bibitem[H{\"o}flich et al.(1996)]{Hoflich1996} H{\"o}flich, P., Wheeler, J.~C., Hines, D.~C., \& Trammell, S.~R.\ 1996, \apj, 459, 307
\bibitem[Howell et al.(2001)]{Howell2001} Howell, D.~A., H{\"o}flich, P., Wang, L., \& Wheeler, J.~C.\ 2001, \apj, 556, 302 
\bibitem[Inserra et al.(2013)]{Inserra2013} Inserra, C., Pastorello, A., Turatto, M., et al.\ 2013, \aap, 555, A142 
\bibitem[Janka et al.(2007)]{Janka2007} Janka, H.-T., Langanke, K., Marek, A., Mart{\'{\i}}nez-Pinedo, G., \& M{\"u}ller, B.\ 2007, \physrep, 442, 38 
\bibitem[Janka(2012)]{Janka2012} Janka, H.-T.\ 2012, Annual Review of Nuclear and Particle Science, 62, 407
\bibitem[Jehin et al.(2005)]{Jehin2005} Jehin, E., O'Brien, K., \& Szeifert, T. 2005, FORS1+2 User's Manual, VLT-MAN-ESO-13100-1543, Issue 78
\bibitem[Andrews et al.(2019)]{Andrews2019} Andrews J.~E., et al., 2019, arXiv, arXiv:1907.01013
\bibitem[Kasen et al.(2006)]{Kasen2006} Kasen, D., Thomas, R.~C., \& Nugent, P.\ 2006, \apj, 651, 366
\bibitem[Kumar et al.(2016)]{Kumar2016} Kumar, B., Pandey, S.~B., Eswaraiah, C., \& Kawabata, K.~S.\ 2016, \mnras, 456, 3157 
\bibitem[Leonard et al.(2000)]{Leonard2000} Leonard, D.~C., Filippenko, A.~V., Barth, A.~J., \& Matheson, T.\ 2000, \apj, 536, 239 
\bibitem[Leonard et al.(2001)]{Leonard2001} Leonard, D.~C., Filippenko, A.~V., Ardila, D.~R., \& Brotherton, M.~S.\ 2001, \apj, 553, 861 
\bibitem[Leonard et al.(2005)]{Leonard2005} Leonard, D.~C., Li, W., Filippenko, A.~V., Foley, R.~J., \& Chornock, R.\ 2005, \apj, 632, 450
\bibitem[Leonard et al.(2006)]{Leonard2006} Leonard, D.~C., Filippenko, A.~V., Ganeshalingam, M., et al.\ 2006, \nat, 440, 505 
\bibitem[Li et al.(2011)]{Li2011} Li, W., Leaman, J., Chornock, R., et al.\ 2011, \mnras, 412, 1441 
\bibitem[Liebend{\"o}rfer et al.(2001)]{Liebendorfer2001} Liebend{\"o}rfer, M., Mezzacappa, A., Thielemann, F.-K., et al.\ 2001, \prd, 63, 103004 
\bibitem[Marek \& Janka(2009)]{Marek2009} Marek, A., \& Janka, H.-T.\ 2009, \apj, 694, 664 
\bibitem[Mauerhan et al.(2017)]{Mauerhan2017} Mauerhan, J.~C., Van Dyk, S.~D., Johansson, J., et al.\ 2017, \apj, 834, 118 
\bibitem[Mauron \& Josselin(2011)]{Mauron2011} Mauron, N., \& Josselin, E.\ 2011, \aap, 526, A156 
\bibitem[Melson et al.(2015)]{Melson2015} Melson, T., Janka, H.-T., \& Marek, A.\ 2015, \apjl, 801, L24 
\bibitem[Meyer et al.(2004)]{Meyer2004} Meyer, M.~J., Zwaan, M.~A., Webster, R.~L., et al.\ 2004, \mnras, 350, 1195 
\bibitem[Nagao et al.(2017)]{Nagao2017} Nagao, T., Maeda, K., \& Tanaka, M.\ 2017, \apj, 847, 111 
\bibitem[Nagao et al.(2018)]{Nagao2018} Nagao, T., Maeda, K., \& Tanaka, M.\ 2018, \apj, 861, 1 
\bibitem[Patat \& Romaniello(2006)]{Patat2006} Patat, F., \& Romaniello, M.\ 2006, \pasp, 118, 146 
\bibitem[Patat et al.(2012)]{Patat2012} Patat, F., H{\"o}flich, P., Baade, D., et al.\ 2012, \aap, 545, A7 
\bibitem[Pursimo et al.(2017)]{Pursimo2017} Pursimo, T., Elias-Rosa, N., Dennefeld, M., et al.\ 2017, The Astronomer's Telegram, 10717
\bibitem[Rampp \& Janka(2000)]{Rampp2000} Rampp, M., \& Janka, H.-T.\ 2000, \apjl, 539, L33
\bibitem[Reilly et al.(2017)]{Reilly2017} Reilly, E., Maund, J.~R., Baade, D., et al.\ 2017, \mnras, 470, 1491 
\bibitem[Sanders et al.(2015)]{Sanders2015} Sanders, N.~E., Soderberg, A.~M., Gezari, S., et al.\ 2015, \apj, 799, 208
\bibitem[Schlafly \& Finkbeiner(2011)]{Schlafly2011} Schlafly, E.~F., \& Finkbeiner, D.~P.\ 2011, \apj, 737, 103 
\bibitem[Serkowski et al.(1975)]{Serkowski1975} Serkowski, K., Mathewson, D.~S., \& Ford, V.~L.\ 1975, \apj, 196, 261 
\bibitem[Sumiyoshi et al.(2005)]{Sumiyoshi2005} Sumiyoshi, K., Yamada, S., Suzuki, H., et al.\ 2005, \apj, 629, 922 
\bibitem[Takiwaki et al.(2012)]{Takiwaki2012} Takiwaki, T., Kotake, K., \& Suwa, Y.\ 2012, \apj, 749, 98 
\bibitem[Tartaglia et al.(2018)]{Tartaglia2018} Tartaglia, L., Sand, D.~J., Valenti, S., et al.\ 2018, \apj, 853, 62 
\bibitem[Thompson et al.(2003)]{Thompson2003} Thompson, T.~A., Burrows, A., \& Pinto, P.~A.\ 2003, \apj, 592, 434 
\bibitem[Trammell et al.(1993)]{Trammell1993} Trammell, S.~R., Hines, D.~C., \& Wheeler, J.~C.\ 1993, \apjl, 414, L21 
\bibitem[Tran et al.(1997)]{Tran1997} Tran, H.~D., Filippenko, A.~V., Schmidt, G.~D., et al.\ 1997, \pasp, 109, 489 
\bibitem[Valenti et al.(2016)]{Valenti2016} Valenti, S., Howell, D.~A., Stritzinger, M.~D., et al.\ 2016, \mnras, 459, 3939 
\bibitem[Valenti et al.(2017)]{Valenti2017} Valenti, S., Tartaglia, L., Sand, D., et al.\ 2017, The Astronomer's Telegram, 10706
\bibitem[Vink{\'o} et al.(2006)]{Vinko2006} Vink{\'o}, J., Tak{\'a}ts, K., S{\'a}rneczky, K., et al.\ 2006, \mnras, 369, 1780 
\bibitem[Wang \& Wheeler(1996)]{Wang1996} Wang, L., \& Wheeler, J.~C.\ 1996, \apjl, 462, L27 
\bibitem[Wang et al.(1997)]{Wang1997} Wang, L., Wheeler, J.~C., \& H{\"o}flich, P.\ 1997, \apjl, 476, L27
\bibitem[Wang et al.(2001)]{Wang2001} Wang, L., Howell, D.~A., H{\"o}flich, P., \& Wheeler, J.~C.\ 2001, \apj, 550, 1030 
\bibitem[Wang et al.(2004)]{Wang2004} Wang, L., Baade, D., H{\"o}flich, P., et al.\ 2004, \apjl, 604, L53
\bibitem[Wang \& Wheeler(2008)]{Wang2008} Wang, L., \& Wheeler, J.~C.\ 2008, \araa, 46, 433 
\end{thebibliography}







\bsp	
\label{lastpage}
\end{document}